\documentclass[twocolumn]{emulateapj}
\usepackage[latin2]{inputenc}
\usepackage{hyperref}

\mathchardef\mhyphen="2D

\newcommand{\bssC}{\textbf{\textsf{C}}}

\newcommand{\R}{\mathcal{R}}
\newcommand{\avg}[1]{\left\langle{#1}\right\rangle}

\newcommand{\hmpc}{\,$h^{-1}$\,Mpc}
\newcommand{\hgpcnosp}{$h^{-1}$\,Gpc}

\newcommand{\ihmpc}{\,$h$\,Mpc$^{-1}$}
\newcommand{\impc}{Mpc$^{-1}$}
\newcommand{\ihmpcnosp}{$h$\,Mpc$^{-1}$}

\newcommand{\pd}{$P_\delta$}

\newcommand{\pln}{$P_{\ln(1+\delta)}$}
\newcommand{\pg}{$P_{G(\delta)}$}

\newcommand{\pdk}{$P_{\delta}(k)$}
\newcommand{\plnk}{$P_{\ln(1+\delta)}(k)$}
\newcommand{\pgk}{$P_{G(\delta)}(k)$}
\newcommand{\pds}{$P_{\delta/\sigma}$}

\newcommand{\plin}{$P_{\rm lin}$}
\newcommand{\pnowig}{$P_{\rm nowig}$}
\newcommand{\camb}{{\scshape camb}}
\newcommand{\cosmicemu}{{\scshape CosmicEmu}}

\newcommand{\sns}{(S/N)$^2$}

\newcommand{\kmin}{k_{\rm min}}
\newcommand{\kmax}{k_{\rm max}}

\newcommand{\balpha}{\mbox{\boldmath$\alpha$}}
\newcommand{\barbalpha}{\mbox{\boldmath$\bar{\alpha}$}}
\newcommand{\boldcdot}{\mbox{\boldmath$\cdot$}}
\newcommand{\bD}{\mbox{\boldmath$D$}}
\newcommand{\lses}{\ln \sigma_8^2}

\chardef\til=`\~

\begin{document}

\title{Rejuvenating the Matter Power Spectrum III:\\
  The Cosmology Sensitivity of Gaussianized Power Spectra}

\author{Mark C. Neyrinck\\
{\rm \small 
Department of Physics and Astronomy, The Johns Hopkins University, Baltimore, MD 21218, USA}
}

\begin{abstract}
It was recently shown that applying a Gaussianizing transform, such as
a logarithm, to the nonlinear matter density field extends the range
of useful applicability of the power spectrum by a factor of a few
smaller.  Such a transform dramatically reduces nonlinearities in both
the covariance and the shape of the power spectrum.  Here, analyzing
Coyote Universe real-space dark matter density fields, we investigate
the consequences of these transforms for cosmological parameter
estimation.  The power spectrum of the log-density provides the
tightest cosmological parameter error bars (marginalized or not),
giving a factor of 2-3 improvement over the conventional power
spectrum in all five parameters tested.  For the tilt, $n_s$, the
improvement reaches a factor of 5.  Similar constraints are achieved
if the log-density power spectrum and conventional power spectrum are
analyzed together.  Rank-order Gaussianization seems just as useful as
a log transform to constrain $n_s$, but not other parameters.
Dividing the overdensity by its dispersion in few-Mpc cells, while it
diagonalizes the covariance matrix, does not seem to help with
parameter constraints.  We also provide a code that emulates these
power spectra over a range of concordance cosmological models.
\end{abstract}

\keywords{cosmology: theory --- cosmology: observations --- large-scale structure of universe --- methods: statistical}

\section{Introduction}
The distribution of matter in the Universe on large scales is
efficiently quantified by the power spectrum of its overdensity
fluctuations.  This is because to a good approximation, the density
field is a Gaussian random field, at early times (as established by
observations of the cosmic microwave background, CMB), or on large
scales at low redshift.  The problem of efficient structure
quantification is interesting information-theoretically, and also
practically, for constraining cosmological parameters.

While the power spectrum of the overdensity
$\delta=(\rho-\bar\rho)/\bar\rho$ is the optimal statistic on the
largest scales, on translinear scales ($0.2\lesssim
k/[$\ihmpcnosp$]\lesssim 0.8$ at redshift $z=0$ in a concordance
$\Lambda$CDM model), the dark-matter density field departs
substantially from Gaussianity, and the power spectrum covariance
matrix develops a significant non-Gaussian component
\citep{mw,szh,t09}.  This leads to a plateau in Fisher information
\citep{rh05,rh06,ns06,leepen}; when measuring a parameter such as the
initial power spectrum amplitude, modes in the translinear regime are
highly correlated, giving little additional constraining power when
analyzed with larger-scale modes.

This signals a failure of the power spectrum to describe the $\delta$
field fully on these scales, and more practically, implies a
substantial reduction in its power to constrain cosmological
parameters.  The number of Fourier modes grows as $k^3$ for a 3D
survey, and so it would be a shame if these smaller-scale modes could
not be used to constrain cosmology.

Methods have been proposed that reduce the covariance on translinear
scales to varying degrees. These include pre-whitening
\citep{ajsh2000}; removing large halos from a survey \citep{ns07}; a
Gaussianizing transform; nonlinear wavelet Wiener filtering
\citep{zhang11}; and dividing $\delta$ by its dispersion in few-Mpc
cells \citep{n11a}.  In this paper, by ``Gaussianizing transform'' we
mean a function applied to pixels in a field (e.g.\ $\delta$ measured
in cells of some resolution) that increases the Gaussianity of 1-point
probability density function (PDF) of this field.  Examples of
Gaussianizing transforms for the cosmological density field include: a
logarithmic $\delta\to\ln(1+\delta)$ transform \citep[]{nss09,seo};
rank-order Gaussianization $\delta\to G(\delta)$, giving an exactly
Gaussian distribution by mapping the 1-point PDF onto a Gaussian of
some width \citep{weinberg,nss09,yuyu}; and a Box-Cox transform
\citep{boxcox,jtk11}, which can be considered a generalization of the
logarithmic transform with parameters tunable to give vanishing
skewness and kurtosis.  A related statistic to the
rank-order-Gaussianized power spectrum is the copula \citep{copula}.

There are reasons to think that $A=\ln(1+\delta)$ would be more
appropriate to analyze than $\delta$. \citet{colesjones} pointed out
theoretically that a lognormal PDF emerges if peculiar velocities are
assumed to grow according to linear theory. Using a
Schr\"{o}dinger-equation framework, \citet{spt} found that the
variance in $A$ is much better-described in tree-level perturbation
theory than the variance in $\delta$, suggesting that $A$ is closer to
linear theory. In a study of discreteness effects, \citet{romeo} also
observed in simulations that the first few moments of $A$ have reduced
fractional variance compared to $\delta$.

Going beyond the inherent statistics of $A$ into parameter-dependence,
\citet{carron} found analytically that for a lognormal density field
$\delta$ whose moments depend on a cosmological parameter, the
underlying Gaussian field $A$ is (often much) more informative about
the parameter than the lognormal field $\delta$.  Also, \citet{jtk11}
found that applying a log transform to a simulated weak-lensing
convergence field allows significantly tighter constraints in a
$\sigma_8$-$\Omega_m$ parameter space.  However, they found that
adding realistic galaxy shape noise in the analysis degrades the
constraints both in the conventional and transformed convergence
fields, reducing the gains from Gaussianizing.

In this paper, we explore the cosmology-constraining power of applying
three of these transformations to the real-space dark-matter
overdensity field: a logarithmic transform, rank-order
Gaussianization, and dividing $\delta$ by its dispersion in cells.
Our analysis ignores the observationally relevant issues of shot noise
and galaxy bias (if the transforms are applied to a galaxy survey),
and redshift-space distortions.  In Paper II \citep{nss11}, we began
the analysis of these issues, to the point that we are confident that
any cosmological-parameter tightening we find in this study will
translate to improvement in a realistic situation as well, although
probably to a smaller degree.

\section{Results}

In Paper I, we showed that Gaussianizing the low-redshift Millennium
simulation \citep{mill} matter-density field seems to restore a linear
shape; here we test this a bit more generally.  \citet{wang11} studied
the scale-dependence of the power spectrum of the log-density, \pln,
in renormalized perturbation theory.  In that paper, in the
perturbatively predicted \pln\ there is a hint of decreased
nonlinearity in its shape compared to the conventional power spectrum
\pd, but the perturbative approach does not reach deeply into
nonlinear scales.

\begin{figure}
  \begin{center}
    \includegraphics[scale=0.4]{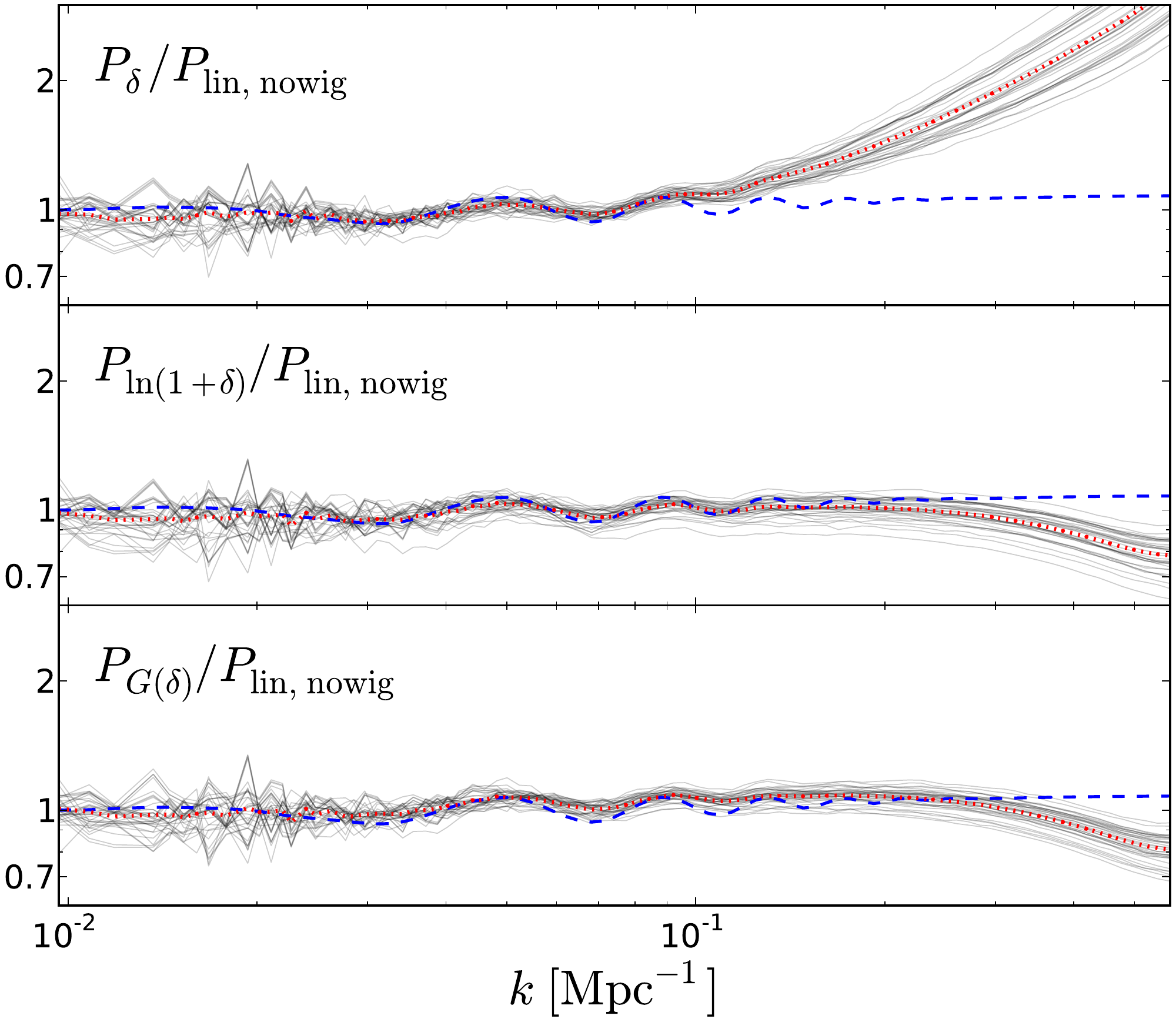}
  \end{center}  
  \caption{Nonlinear transfer functions of 37 Coyote Universe
    simulations spanning a space of five cosmological parameters.  The
    denominators in each panel are no-wiggle linear power spectra,
    without considering effects from the NGP pixel window function.
    The pixel window function causes the small-scale downturns in the
    bottom two panels (and attenuates \pd\ as well).  The dashed blue
    curves show the ratio of \plin\ to \pnowig.  The
    dotted red curves are averages from the 37 simulations.}
  \label{fig:compline}
\end{figure}

Fig.\ \ref{fig:compline} shows, for each of 37 high-resolution
simulations in the Coyote Universe \citep{coyote1,coyote2,coyote3}
suite, ratios of \pd, \pln\ and \pg\ (the power spectrum of the
rank-order-Gaussianized field $G(\delta)$, here mapping $\delta$ onto
a Gaussian of unit variance), to the no-wiggle linear power spectrum
\pnowig\ \citep{ehu}, using that simulation's cosmology.

Each curve is, essentially, a nonlinear transfer function, for a
slightly different cosmology.  There are important subtleties, though.

First, the power spectra are not divided by the power spectrum of the
actual initial conditions, but the ensemble-average linear power
spectrum (and a no-wiggle version of it, at that).  Thus,
cosmic-variance noise is present.  However, we dampen this noise, by
showing, for each density field, the average of the power spectra
after applying 52 (up to the first harmonic in each direction)
sinusoidal weightings \citep[][HRS]{hrs}.

Second, only the simulation power spectra (the numerator, but not the
denominator) are attenuated by nearest-grid-point (NGP) pixel window
functions.  We plot each curve to its Nyquist wavenumber, where the
attenuation is substantial.  We did not correct for the pixel window
function in the measurements because for \pln\ and \pg, changing the
resolution does more than simply introduce a small-scale attenuation,
for example changing the large-scale amplitude.  \pln\ and \pg\ do
lack the nonlinear upward ramp at $k\gtrsim 0.1$ \impc\ present in
\pd, but starting at $k\approx 0.3$ \impc, they turn down compared to
\pnowig.  Compare this to Fig.\ 3 of Paper I, in which the denominator
is the power spectrum of the simulation's exact initial conditions,
including NGP attenuation (possible because of the much higher mass
resolution in the Millennium simulation).  In Fig.\ 3 of Paper I, the
ratios do not depart substantially from unity.  Thus we mainly ascribe
the downturns in Fig.\ \ref{fig:compline} to the resolution-dependent
NGP pixel window function.  The \pd\ curves would also be higher than
plotted at high $k$ without the NGP attenuation.

Finally, \pln\ and \pg, generally biased on large scales compared to
\pd, are multiplied by a factor to line up with \pd\ in the
smallest-$k$ bin shown.  For \pg, this process is equivalent to
setting the variance of the Gaussian onto which $\delta$ is mapped so
that the large-scale amplitude of \pg\ is the same as for \pd.  In all
further analysis below, for simplicity, we use a Gaussian with
variance 1.

We rank-order-Gaussianize $\delta$ before the HRS weightings are
applied, rather than Gaussianizing each weighted density field
separately. Arguably, it would be fairest to Gaussianize each
seperately, since the 1-point PDF of each weighted density field will
slightly fluctuate from the PDF of the unweighted field. In real
observations, one would be confronted with this fluctuated PDF, not
the PDF from a larger sample.  The main reason we Gaussianize before
weighting is that practicically, it is nontrivial to Gaussianize the
weighted density field, which has essentially a highly nonuniform
(sinusoidal) selection function. Our procedure likely slightly
overestimates the covariance from Gaussianizing the weighted density
fields seperately, since this would equalize the variances in each.
This would likely have similar covariance-killing effects as we found
dividing \pd\ by the density variance to have \citep{n11a}.

Each simulation, analyzed at redshift $z=0$, has a different set of
cosmological parameters, each a plausible (given current observational
constraints) concordance cosmological model.  The simulations occupy
an orthogonal-array-Latin-hypercube in the five-dimensional parameter
space $\omega_m=\Omega_mh^2\in[0.12,0.155]$,
$\omega_b=\Omega_bh^2\in[0.0215,0.0235]$, $n_s\in[0.85,1.05]$,
$w\in[-1.3,-0.7]$, $\sigma_8\in[0.61,0.9]$.  The remaining
cosmological parameters, e.g.\ $h$, are set to match the tight CMB
constraint on the ratio of the last-scattering-surface distance to the
sound-horizon scale.  The 1024$^3$-particle simulations have box size
1300 Mpc, fixed in Mpc (not \hmpc) to roughly line up
baryon-acoustic-oscillation (BAO) features in $k$ among different
cosmologies.  Their resolution is sufficient for power-spectrum
measurements accurate at sub-percent level at scales down to
$k=1$\ihmpc.

All the results in this paper use $256^3$ grids, a resolution at which
shot noise is negligible even for \pln\ and \pg.  We do not push to
smaller scales because even at this resolution, 25 of the simulations
have handfuls of cells with zero particles.  Among these 25, the
median number of zero-particle cells is 19, with maximum 341, still
$\ll 256^3$.  To apply the log transform, we set the effective number
of particles $N_{\rm eff}(N=0)=1/2$ in zero-particle cells, i.e.\ as
though there were half a particle in the cell.  This equalizes the
distance in log-density between cells with 0 and 1 particle, and 1 and
2 particles.  We experimented with changing $N_{\rm eff}(N=0)$ by
factors of two up and down (to 1/4 and 1).  Unsurprisingly, this had
little effect; in the worst-case simulation with 341 zero-particle
cells, \pln\ changed by at most (over all $k$) 0.03\%, with typical
changes $\sim$0.01\%.

The level of fluctuation in the nonlinear transfer function looks
substantially greater for \pd\ than for \pln\ and \pg\ at large $k$,
but around 0.1 \impc, there is not much difference.  Some of the
small-scale fluctuation could be from slight cosmology-dependent
inaccuracies in the no-wiggle transfer function.  Note that the dashed
curve, showing the ratio of \plin, from \camb\ \citep{camb}, to
\pnowig, departs slightly from 1 at large $k$.

By eye, the BAO are of similar amplitude in \pln\ and \pg\ as in \pd.
This is not surprising; Gaussianizing does not undo the bulk motions
that erase small-scale BAO wiggles in \pd.  One difference, though, is
that in \pd, the smallest-scale wiggles sit atop the start of the
nonlinear ramp, which suggests that their detection may have to
compete with a shot-noise-like (on translinear scales) one-halo term.
Variance in this term can be seen as the source of the translinear
covariance \citep{ns06}.  Note that in all panels, the BAO are likely
a bit damped or smeared from the averaging over sinusoidal weightings,
but to a lesser degree than the results below, which use up to the
second harmonic in weightings.  This dampening should affect all
panels equally, and the scales of the weightings (just beyond the left
edge of the plot) are $\sim 5$ times larger than BAO scales, so these
effects are probably small.  However, we leave a thorough quantitative
analysis of BAO detection in Gaussianized power spectra to future
work.

\subsection{Covariance Matrices}
We estimate the cumulative Fisher information \citep{fisher,tth,ns07} in
parameters $\alpha$ and $\beta$ over a range of bin indices $i\in\R$
as
\begin{equation}
  F_{\alpha\beta}(\R) =
  \sum_{i,j\in \R} \frac{\partial\ln P_i}{\partial\alpha}(\bssC_\R^{-1})_{ij} 
 \frac{\partial\ln P_j}{\partial\beta},
  \label{eqn:inforange}
\end{equation}
where $\bssC_\R$ is the square submatrix of the power-spectrum
covariance matrix $\bssC$ with both indices ranging over $\R$.
$C_{ij}\equiv\avg{\Delta\ln P_i\Delta\ln P_j}=\avg{\Delta P_i\Delta
  P_j}/\avg{P_i}\avg{P_j}$.  The inverse of $F_{\alpha\beta}$ then
gives the parameter covariance matrix.

In the bin range $\R$, $\kmin$ is the lowest $k$ not directly
modulated by the sinusoidal weightings (beyond $\sqrt{3}$ times the
wavenumber of the second harmonic), i.e.\ $\kmin=2\pi/(1300~ {\rm
  Mpc})\times2\sqrt{3}=0.017~{\rm Mpc}^{-1}$. We investigate
constraints on parameters as $\kmax$ varies, up to the Nyquist
frequency, $\sim 0.6~{\rm Mpc}^{-1}$.  The bins of $\kmax$ vary by
factors of $\sqrt[16]2$ (approximately, since each bin's $\kmax$ is
the mean $\kmax$ in the bin).

The power-spectrum covariance matrices are measured from the Coyote
Universe suite, as in \citet{n11a}.  We used the HRS
sinusoidal-weightings method, going up to the second harmonic to get
248 different power spectra from each simulation. This gave an
estimate of the covariance in $\ln P$ from each simulation.  We then
formed an average covariance matrix across the simulations, to reduce
noise. We averaged the covariances of $\ln P$ instead of $P$ for
numerical stability across cosmologies; e.g.\ in linear theory, the
covariance in $\ln P$ does not depend on the power spectrum.

Although using up to second-order weightings and then averaging
together the covariance estimates among simulations beats down the
noise in the covariance matrix substantially, the noise persists at a
level that likely somewhat biases our results. This is because noise
in a matrix that is inverted generally biases the inverse
\citep{hss07}. Ideally, we would correct for the noise as suggested in
that paper, but a necessary ingredient, the number of independent
samples used for the covariance estimate, is not a straightforward
quantity in the HRS weightings method. Each simulation gives 248 power
spectra of overlapping subsamples; especially at low wavenumber, these
power spectra are not necessarily independent.  But this abundance of
perhaps-redundant power spectra has the advantage of reducing the
noise to a level low enough that e.g.\ it always provided naively
invertible covariance matrices.

\begin{figure}
  \begin{center}
    \includegraphics[scale=0.4]{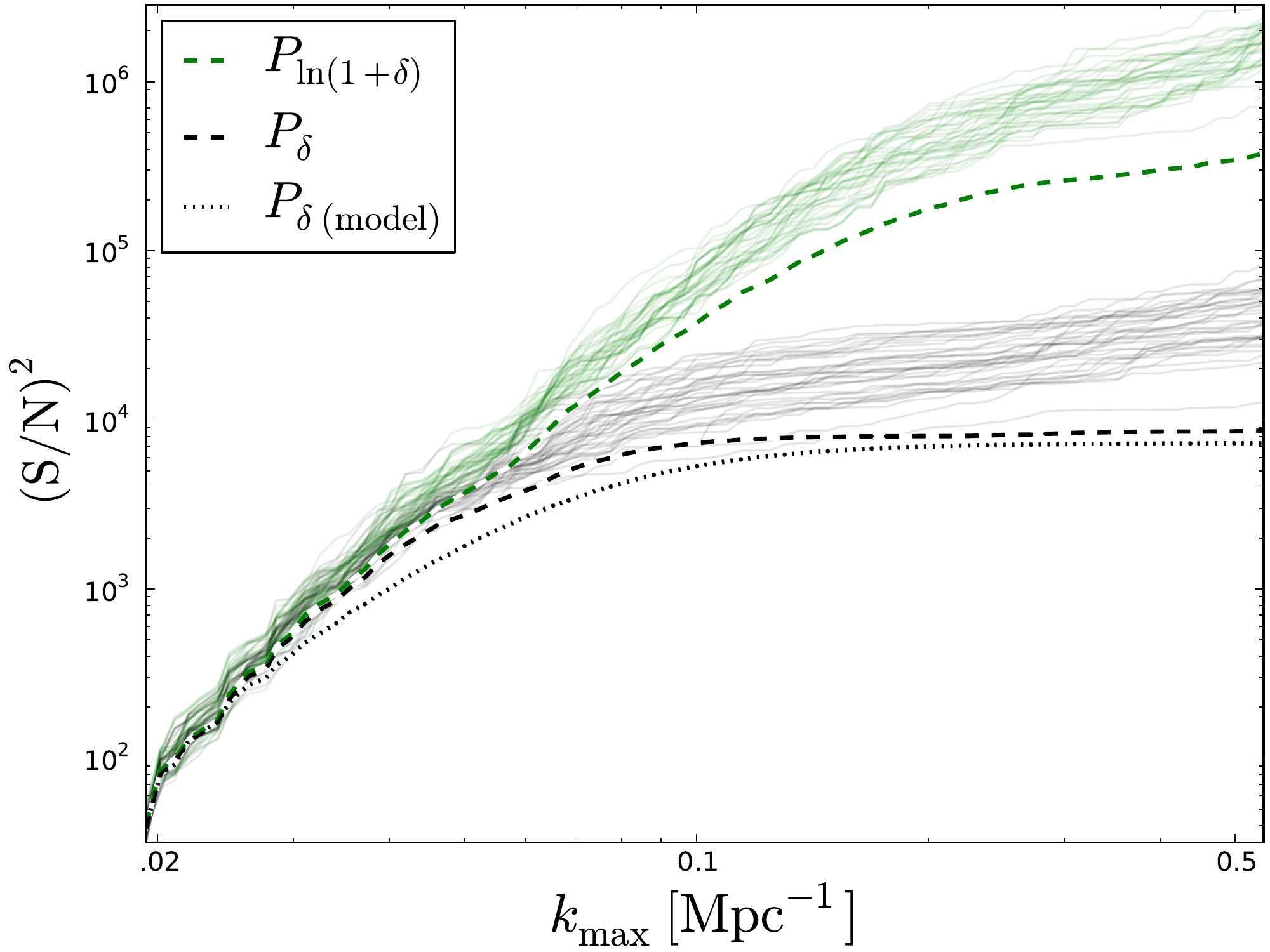}
  \end{center}  
  \caption{The biasing effect of noise in the covariance matrix on the
    Fisher signal-to-noise ratio \sns.  The faint curves show \sns,
    for both \pd\ (black) and \pln\ (green), for each of the 37
    simulations. The bold dashed curves show \sns, using the average
    covariance matrix used for the results below. \sns\ for \pd\ in
    the absence of noise may be estimated using the noise-free (but
    approximate) model of \citet{n11a} (dotted black). The proximity of
    the dashed and dotted black curves suggests a low level of
    residual bias in \sns\ after averaging.}
  
  \label{fig:infonoise}
\end{figure}

Fig.\ \ref{fig:infonoise} illustrates the effect of noise on the
signal-to-noise \sns, indicative of the (inverse) effect on parameter
constraints, as well. Roughly, \sns\ gives the number of statistically
independent modes in the box, $\propto \kmax^3$ for a Gaussian
field. It is measured by setting the derivative factors to unity in
Eq.\ (\ref{eqn:inforange}).  The faint curves show \sns\, from each
simulation; the bold dashed curves show \sns\ using the averaged
covariance, which we use for the results below (proportional to
$1/\sqrt{(S/N)^2}$). Note that the scatter in the faint curves is not
just from ordinary cosmic variance, but (predominantly) from the
scatter in cosmological parameters among the simulations.

Fig.\ \ref{fig:infonoise} shows that much bias in the Fisher matrix is
eliminated by going from a single simulation to an average over 37. To
assess the level of residual bias in \sns\ after the averaging, we
also show an estimate of the \sns\ for \pd\ using a noise-free
approximate covariance matrix. We use the model of \citet{n11a}, in
which the covariance on translinear scales comes from
scale-independent multiplicative fluctuations. Its only ingredient is
the pixel-density variance, measurable with negligible noise. We
averaged together the model covariance matrices in the same way as the
full, measured ones. Comparing \sns\ for \pd\ to this model suggests
that the residual bias is small. Importantly, the bias is also likely
at the same level for all four power spectra investigated, so it
probably does not affect our conclusions. Still, we keep in mind that
our parameter constraints in all cases are likely slightly optimistic.

\begin{figure}
  \begin{center}
    \includegraphics[scale=0.4]{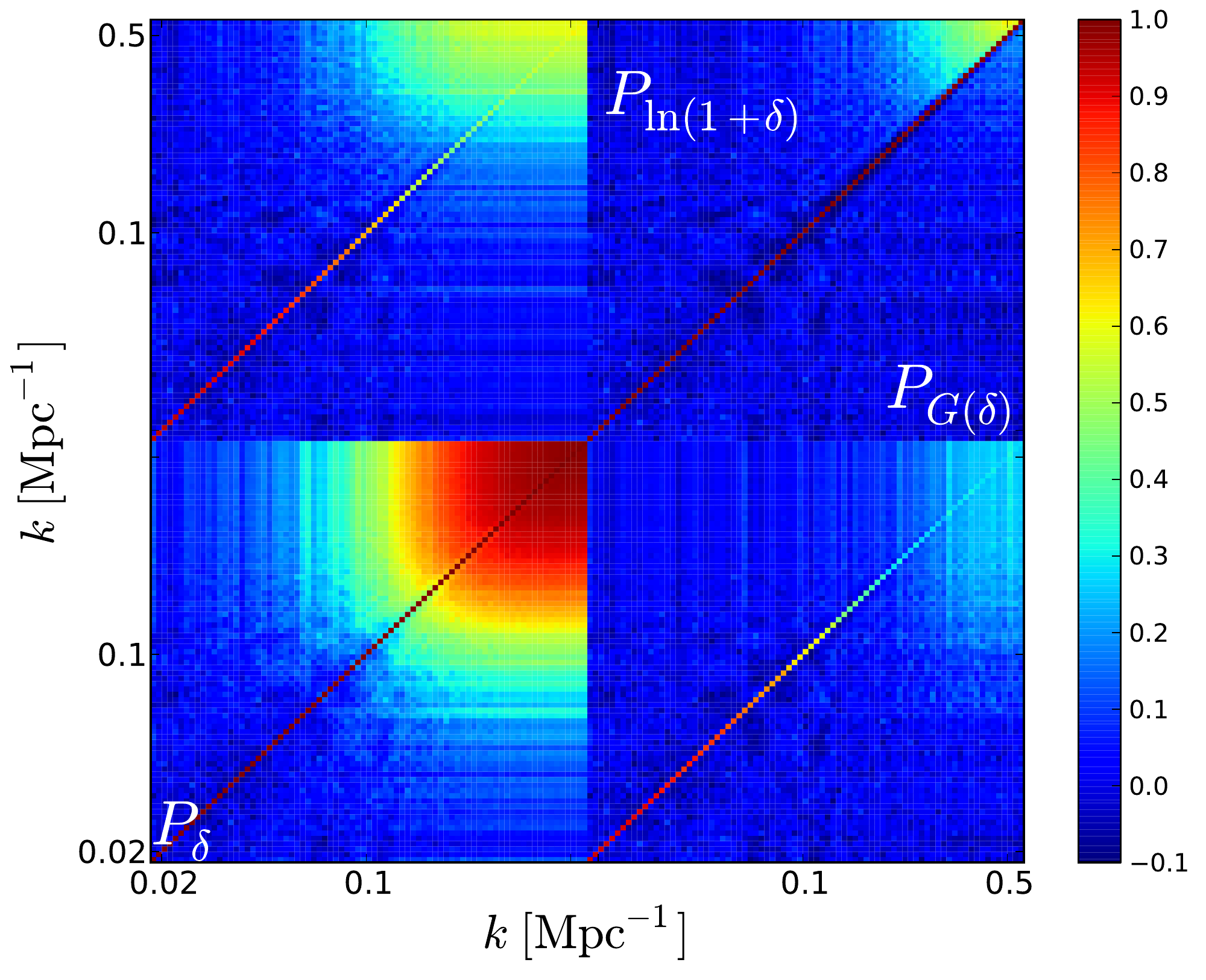}
  \end{center}  
  \caption{Joint correlation matrices of (\pdk\ and \plnk), and
    (\pdk\ and \pgk).  The lower-left square is the correlation matrix
    of \pd\ alone.  The upper-right square is the the correlation
    matrix of \pln\ alone (upper-left corner) and \pg\ alone
    (lower-right corner).  The upper-left square is the
    cross-correlation between bins of \pd\ and \pln; the lower-right
    square is the same for \pd\ and \pg.  The cross-correlation
    diagonals appear in Fig.\ \ref{fig:diagcorr}.}
  \label{fig:triplecorr}
\end{figure}

Fig.\ \ref{fig:triplecorr} shows the joint correlation matrix,
$C_{ij}/\sqrt{C_{ii}C_{jj}}$, for both (\pdk, \plnk) and (\pdk, \pgk).
We need their cross-covariance matrices below when we analyze pairs of
power spectra together.

\begin{figure}
  \begin{center}
    \includegraphics[scale=0.3]{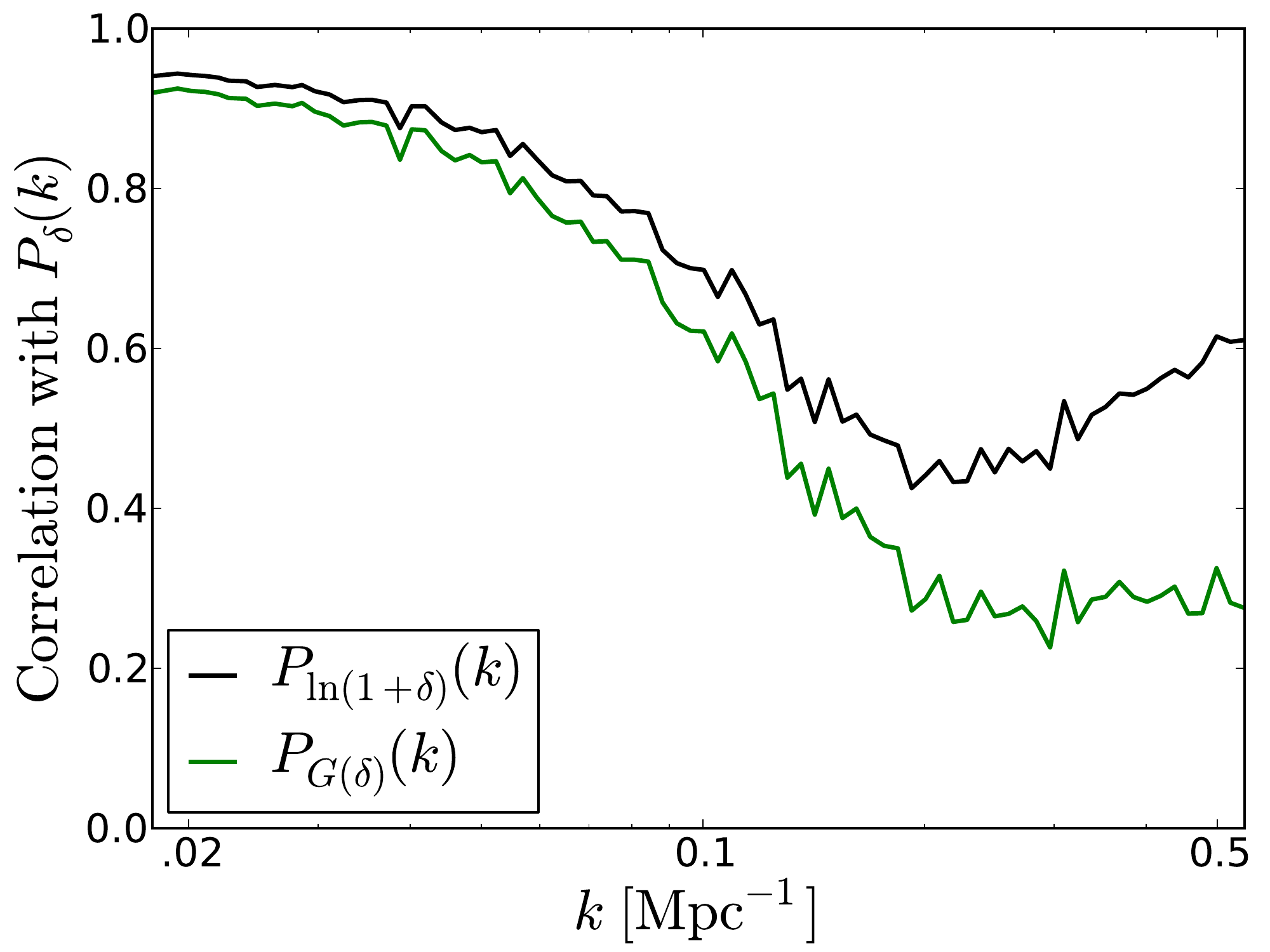}
  \end{center}  
  \caption{Cross-correlations, measured at the same $k$ bin, between
    \pdk\ and \plnk, and \pdk\ and \pgk.  These diagonals do dominate
    their respective cross-correlation matrices at low $k$.  At high
    $k$ though, the curves become small, and are increasingly
    difficult to find amid the off-diagonal covariance.}
  \label{fig:diagcorr}
\end{figure}

Fig.\ \ref{fig:diagcorr} shows the cross-correlation diagonals,
i.e.\ the correlations between \pdk\ and \plnk, and \pdk\ and \pgk, in
the same $k$ bins.  These cross-correlations are related to the
nonlinear propagators \citep{crocce} of each power spectrum, since the
nonlinear propagator of \pd\ is unity on linear scales.  The nonlinear
propagator quantifies the memory of the particular Fourier phases and
amplitudes of the initial conditions, as a function of $k$.  For \pg,
this function dips down to $\sim 0.2$ on translinear scales.

The way we interpret this figure is that all three power spectra have
similar nonlinear propagators, consistent with the findings of
\citet{wang11}.  However, they each have different mode-coupling, or,
roughly speaking, ``one-halo'' terms, correlated to each other only at
the $\sim$20\% level.

\subsection{Derivative Terms}

Here we describe how we estimate the derivative terms relevant for
Fisher analysis, i.e.\ the vector $D_i(k)\equiv \partial\ln
P(k)/\partial\alpha_i$.  The Coyote Universe simulations were set up
to span the 5-parameter space in an optimal way.  \citet{coyote3}
employed sophisticated techniques such as principal-component analysis
and Gaussian-process modeling to produce a precision power-spectrum
emulator, \cosmicemu.  They also used many lower-resolution
simulations to analyze larger-scale modes, that we cannot use here
because the Gaussianized power spectra are more sensitive to particle
discreteness.

Our simpler approach is cruder, but acceptable for our purposes.  In
each $k$ bin, we model the fluctuations away from the mean power
spectrum of all simulations as a linear combination of contributions
from each parameter fluctuation, i.e.\
\begin{equation}
  \ln P^{\balpha}(k) = \avg{\ln P(k)}+\bD(k) \boldcdot (\balpha-\barbalpha),
  \label{eqn:matrixfluc}
\end{equation}
where $\balpha$ is a vector in the space of the five parameters ($\ln
\omega_m$, $\ln \omega_b$, $n_s$, $w$, $\ln\sigma_8^2$).  $\barbalpha$
is the mean of $\balpha$ over all simulations. Linear algebra yields
an estimate of the derivative terms $\bD(k)$ from a quintet of only
five simulations, but it is unusably noisy, the signal swamped by
cosmic variance in each simulation.

We enhance the signal in two ways.  First, in each simulation, we use
not the raw $P(k)$, but the ensemble-average of the power spectra of
the density field after applying the 248 sinusoidal HRS weightings.
This particularly squashes fluctuations away from the mean at small
$k$, while preserving the overall shape.  However, the window
functions of the weightings likely convolve neighboring $k$ modes
together somewhat.  In particular, this probably dampens BAO wiggles a
bit for all power spectra.

\begin{figure}
  \begin{center}
    \includegraphics[scale=0.4]{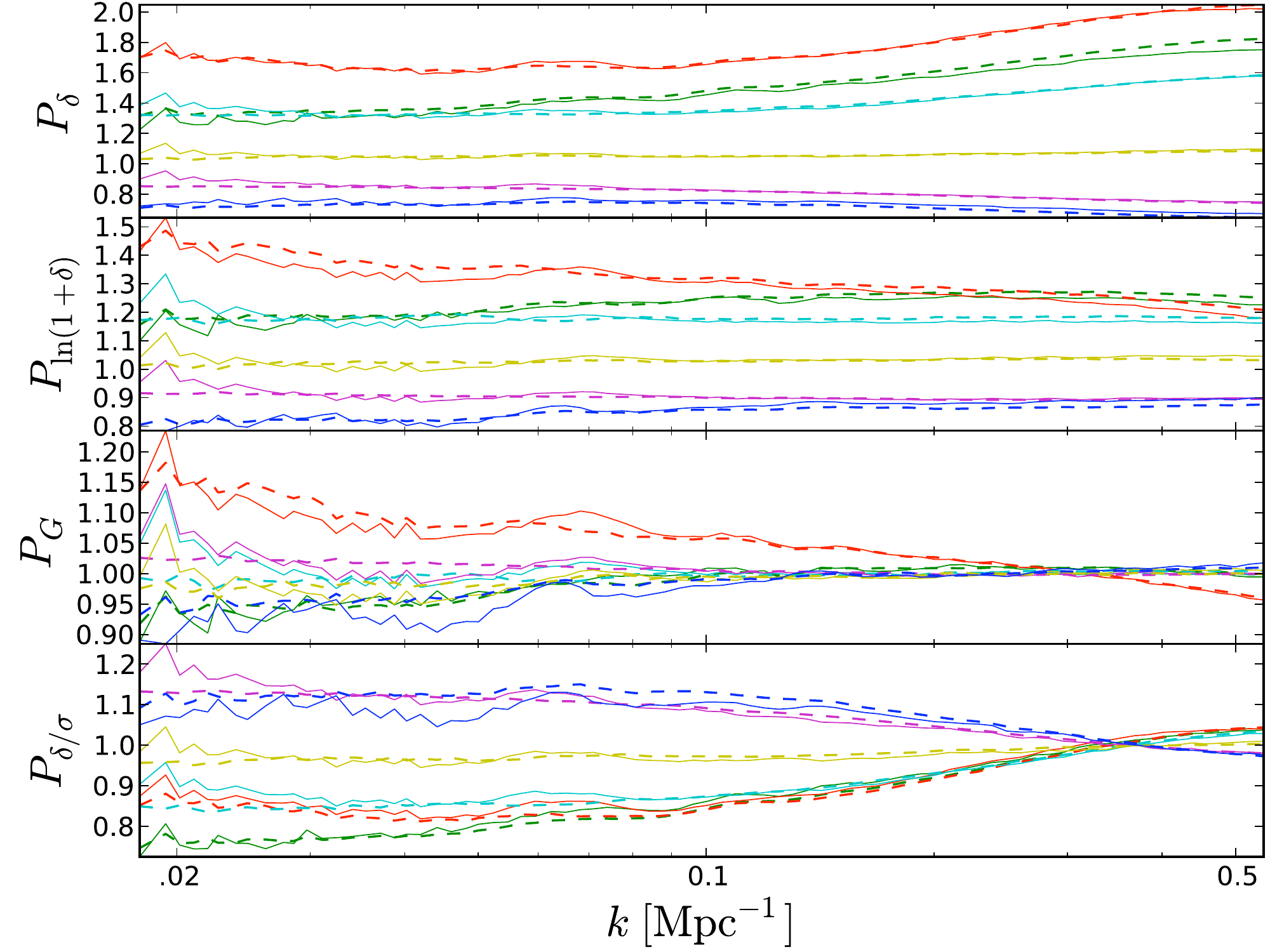}
  \end{center}  
  \caption{Power spectra from Coyote Universe simulations 1-6 (solid),
    together with the predictions from Eq.\ (\ref{eqn:matrixfluc}).
    Cosmic-variance noise in each case has been dampened substantially
    by averaging over 248 HRS sinusoidal weightings.  Only the
    fluctuation, second term on the RHS of Eq.\ (\ref{eqn:matrixfluc})
    is shown, i.e.\ the geometric mean among the simulations is
    divided out.}
  \label{fig:accuracy}
\end{figure}

Second, we estimate $\bD(k)$ by finding the median in each $k$ bin
among many estimates of $\bD(k)$, each measured from a quintet sampled
from the 37 simulations.  Although the results visually converge with
only $\sim 10^3$ quintets, we use $\sim 3\times 10^6$ quintets
(limited by memory), looping through all quartets of simulations
twice, each time choosing a random simulation to complete the quintet.

Fig.\ \ref{fig:accuracy} shows power spectra from simulations 1-6,
compared with their estimates from Eq.\ (\ref{eqn:matrixfluc}).  Each
power spectrum is divided by its geometric mean among the 37
simulations.  Typically, the accuracy is at the few-percent level,
with occasional deviations up to 10\%.  However, these larger
deviations could be from cosmic variance in the particular simulation.
We provide an emulator of these four power 
spectra\footnote{\url{http://skysrv.pha.jhu.edu/~neyrinck/CosmicEmuLog/}}, 
but caution prospective users to note the above caveats.

\begin{figure*}
  \begin{minipage}{175mm}
    \begin{center}
      \includegraphics[scale=0.6]{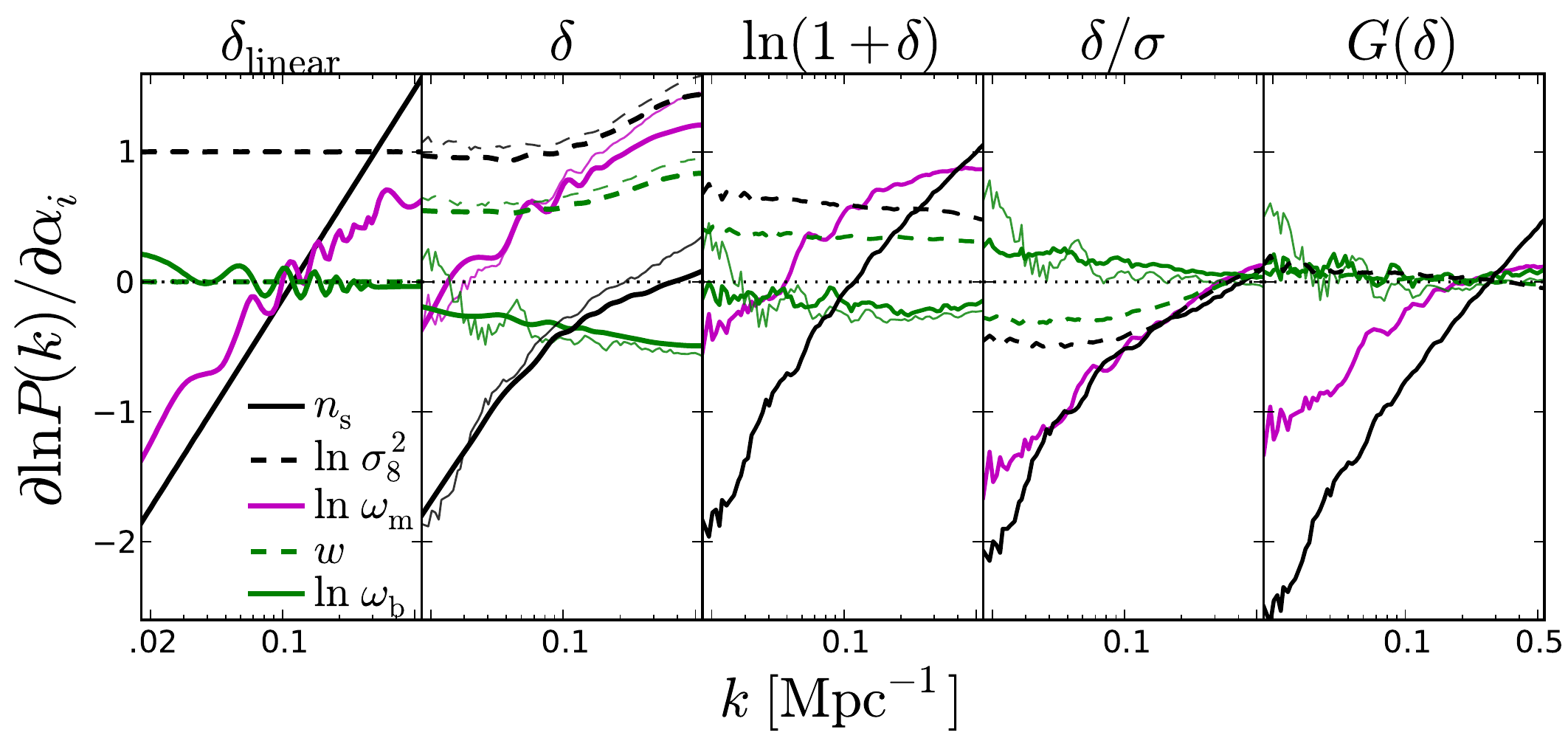}
    \end{center}  
    \caption{Derivative terms $D_i(k)=\partial\ln
      P(k)/\partial\alpha_i$, quantifying the sensitivity of each
      power spectrum to the five explored cosmological parameters.
      The left panel is computed from \camb.  The second, $\delta$
      panel shows $D_i(k)$ estimated as described in the text (thin),
      and from \cosmicemu\ (bold).  For $\delta$, we use our own
      estimates of $D_i(k)$ for further calculations (except for
      $D_{\ln \omega_b}$, for which we use the \cosmicemu\ estimate).
      The (bold) curves in the rightmost three panels show our
      estimates of $D_i(k)$, that we use for further results.  For
      $\omega_b$, a correction of $D_{\ln\omega_b}^{\mbox{\tiny
          \cosmicemu}}(k)-D_{\ln\omega_b}(k)$, measured for \pd, has
      been applied.  The original $\omega_b$ measurements, without the
      correction, appear as the thin curves. Note that slight
      differences between curves are unlikely to be statistically
      significant, given the occasionally visible level of noise.
    }
    \label{fig:dlogs}
  \end{minipage}
\end{figure*}

Fig.\ \ref{fig:dlogs} shows $\bD(k)$ estimated in this way for \plin\
(estimated using \camb), \pd, \pln, \pg, and \pds\ (the power spectrum
of $\delta/\sigma_{\rm cell}$, the $\delta$ divided by its dispersion
in cells).  Except for \plin, and particularly in the latter three
cases, these curves depend on the resolution used (a 256$^3$ grid
here).  However, the general trends here should hold for different
resolutions.  For \pd, generally the results match those measured from
the \cosmicemu\ emulator well.  However, there is a large amount of
noise that produces a discrepancy at small $k$ for
$D_{\ln\omega_b}(k)$.  This is by far the parameter with the smallest
range explored; thus, it is not surprising that our rather crude
method, without further low-resolution simulations for low-$k$ modes,
is not quite adequate to explore it.  For just the parameter
$\omega_b$, we use $D_{\ln\omega_b}(k)$ evaluated from \cosmicemu\ for
\pd.  To all other power spectra, we add the correction
$D_{\ln\omega_b}^{\mbox{\scriptsize \cosmicemu}}(k)-
D_{\ln\omega_b}(k)$.  This seems to improve the accuracy (or at least
decrease the noise), but we caution that our results for $\omega_b$
are less accurate than for other parameters.

A few aspects of Fig.\ \ref{fig:dlogs} should be pointed out.  The
parameter with the most straightforward connection to the
power-spectrum shape is $n_s$, the tilt.  The shape of $D_{n_s}(k)$ for
\pln\ and \pg\ is nearly as straight as that for the linear power
spectrum.  In contrast, for \pd\ and \pds, $D_{n_s}(k)$ bends much more
substantially at the onset of nonlinearity, indicating decreased
sensitivity on small scales.  This supports the claim that
Gaussianization and the log-transform dramatically reduce
nonlinearities in the power-spectrum shape.

A parameter whose simplicity is obscured by unfortunate notation is
$\lses$, which we investigate instead of $\sigma_8$ because for all
$\sigma_8$, $D_{\lses}(k)=1$ in linear theory.  Reassuringly, like
\cosmicemu, for this derivative we obtain about 1 on linear scales for
\pd, as does \cosmicemu.  At this cell resolution for \pln, this
derivative term is decreased to about 0.7, indicating decreased
sensitivity of the mean power spectrum at each wavenumber.  It is
smaller than for \pd\ because the log transform generally decreases
the large-scale amplitude, by a factor of about $\exp(\sigma_{\rm
  cell}^2)$ (Paper I).

In Fig.\ \ref{fig:dlogs}, all \pg\ and \pds\ derivative terms are tied
together at zero at small scales.  This is because for each, the
variance is unity in $1300/256$ Mpc $\approx$ 4\hmpc\ cells.  The derivative
terms are generally smaller in absolute value for these power spectra,
which translates into poorer parameter constraints below than
for \pd\ and \pln, with the exception of the parameter $n_s$.
Curiously, at this resolution, $D_{\lses}(k)$ is of comparable
absolute magnitude for \pds\ and \pln\ for small $k$.  Naively, one
might expect all information about the amplitude to be destroyed in
\pds, in which one divides the power spectrum by the variance in
grid-cell densities (here, of 4\hmpc\ size, but this holds to some
degree for 8\hmpc\ cells).  However, recall that the amplitude
$\sigma_8^2$ is the variance in 8\hmpc\ volumes in the linearly, not
nonlinearly, evolved density field; the nonzero $D_{\lses}(k)$ at
small scales is apparently from the rise in the nonlinear power
spectrum in \pd\ and \pds.

\subsection{Error ellipses}

\begin{figure*}
  \begin{minipage}{175mm}
    \begin{center}
      \includegraphics[scale=0.57]{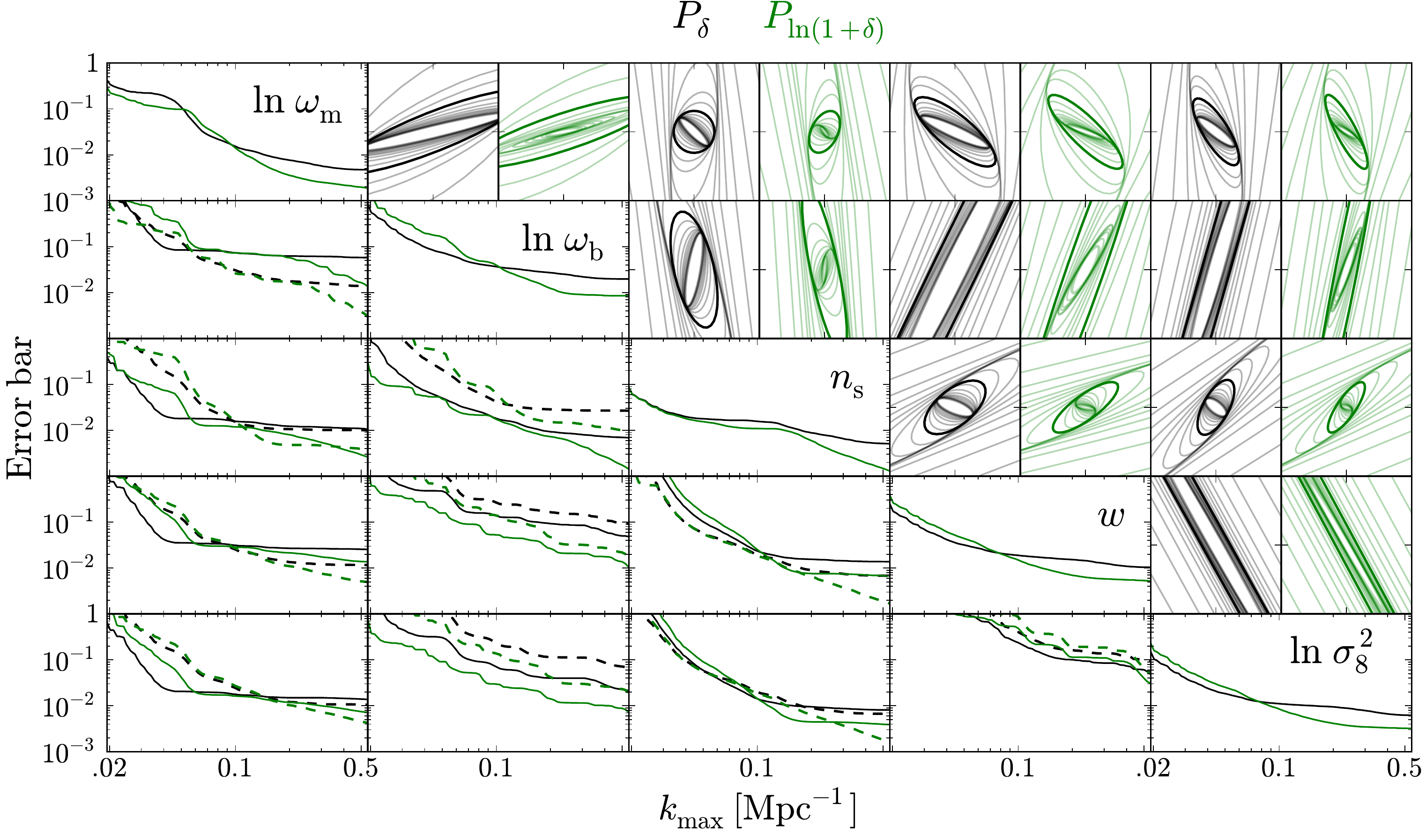}
    \end{center}  
    \caption{Error-bar widths and error ellipses for \pd\ (black) and
      \pln\ (green), in the Coyote-Universe space of five cosmological
      parameters.  Along the diagonal, the curves are unmarginalized
      (half-)error bars over single parameters, holding all else
      fixed.  Below the diagonal, pairs of parameters are considered;
      errors are shown in each parameter, marginalizing over the
      other.  Solid curves correspond to the parameter of a panel's
      row; dashed curves correspond to the column. Above the diagonal,
      error ellipses contract as $\kmax$ increases.  The ellipses, in
      square panels of side half-length 0.5, are centered at the means
      of their parameters. An ellipse is drawn for each bin of $\kmax$
      running from 0.02 to 0.6 Mpc$^{-1}$, log-spaced by a factor of
      $\sqrt[4]{2}$ (four times sparser than the bins constituting the
      curves).  The bold ellipses are at $\kmax=0.1$. Outside,
      analyzing only large scales, \pd\ and \pln\ both give similar
      constraints.  Inside the bold ellipses, nonlinear scales, where
      \pln\ excels, are included, up to the innermost ellipse with
      $\kmax \approx 0.6$ Mpc$^{-1}$.  }
    \label{fig:ellipses_pplog}
  \end{minipage}
\end{figure*}

Fig.\ \ref{fig:ellipses_pplog} shows error bars over the set of five
cosmological parameters, for \pd\ and \pln.  The effective volume for
these results is (1.3 Gpc)$^3/2\approx 1.1$ Gpc$^3\approx .5$
(\hgpcnosp)$^3$.  The factor of two is from the sinusoidal weightings
used for the covariance matrices, which effectively halve the volume.
Along the diagonal, the curves are unmarginalized error bars over
single parameters, holding all else fixed.  Off the diagonal, we
examine error bars allowing sets of two parameters to vary at a
time. The upper plots show how error ellipses contract as $k_{\rm
  max}$ increases, while the lower plots show how marginalized error
bars shrink.

\begin{figure*}
  \begin{minipage}{175mm}
    \begin{center}
      \includegraphics[scale=0.57]{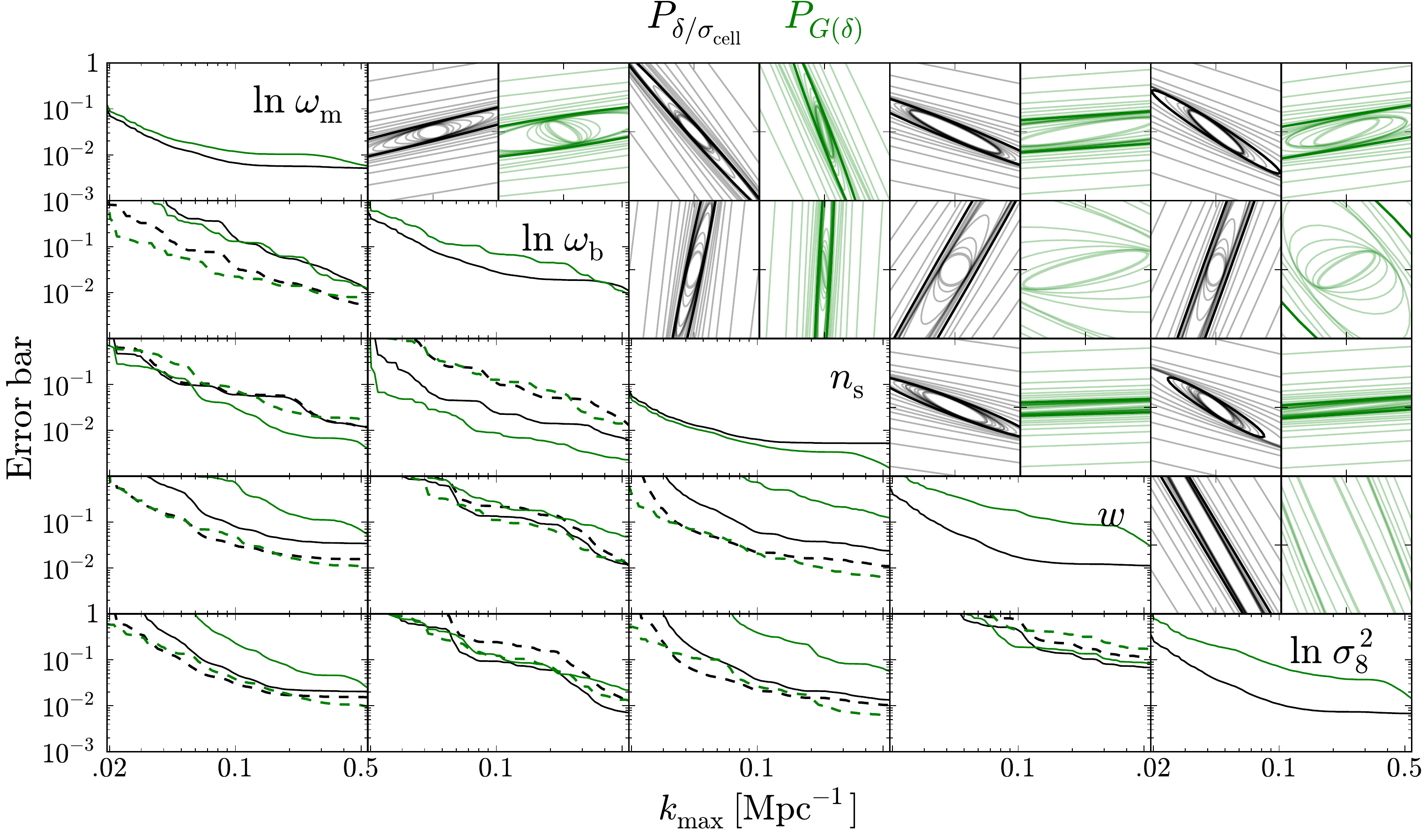}
    \end{center}  
    \caption{Same as Fig.\ \ref{fig:ellipses_pplog}, for \pds\ and \pg.}
    \label{fig:ellipses_pvpg}
  \end{minipage}
\end{figure*}

\begin{figure*}
  \begin{minipage}{175mm}
    \begin{center}
      \includegraphics[scale=0.57]{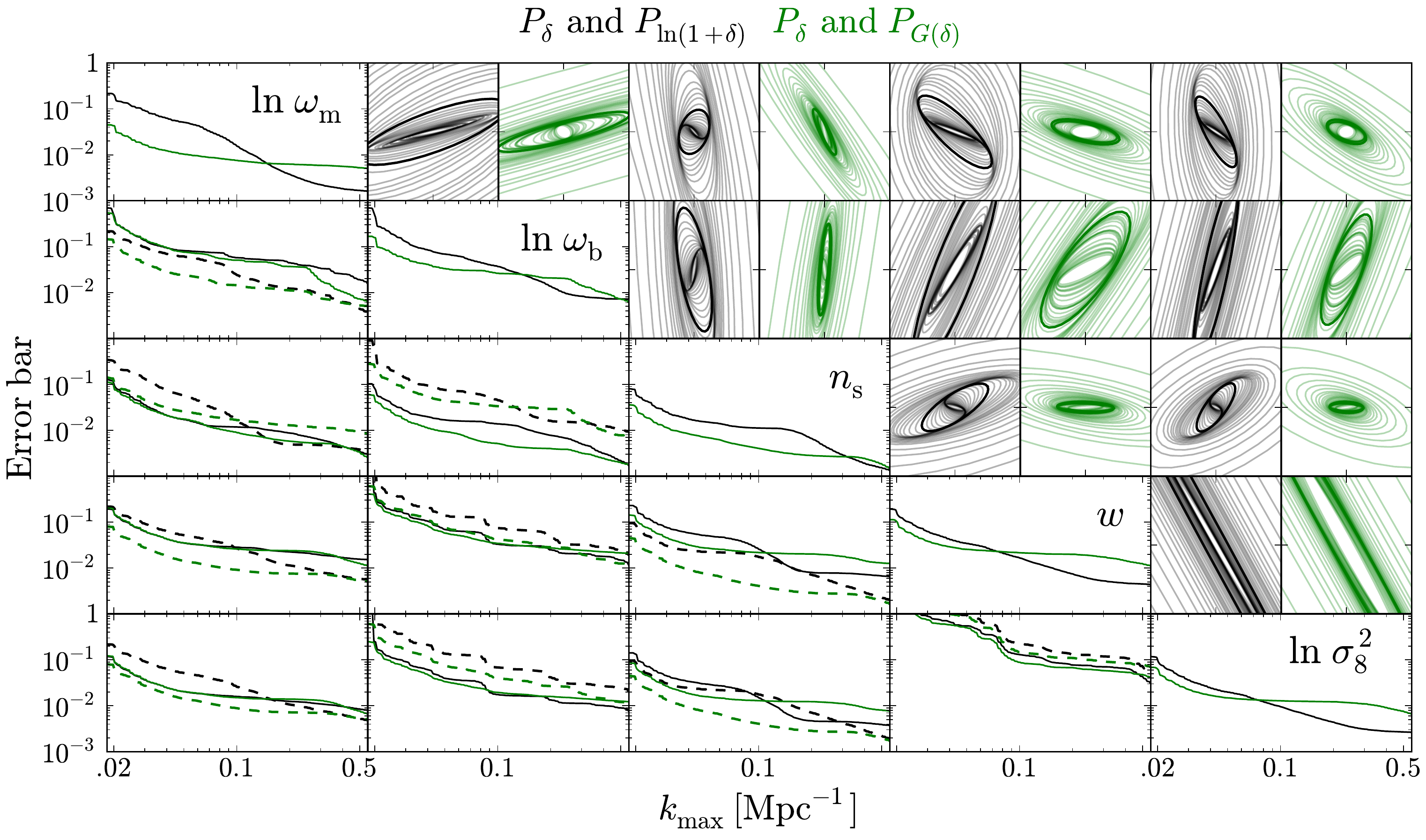}
    \end{center}  
    \caption{Same as Fig.\ \ref{fig:ellipses_pplog}, except we show the
      constraints from analyzing both \pd\ and \pln, and \pd\ and \pg,
      together.  For the combination of \pd\ and \pln, the constraints
      are about the minimum of the constraints from each power
      spectrum separately.  For the \pd-\pg\ combination, though,
      there is some improvement when analyzing both together.}
    \label{fig:ellipses_combined}
  \end{minipage}
\end{figure*}

Constraints obtained analyzing \pln\ are substantially smaller than
for \pd, for all parameters, typically by a factor of 2 or 3 if the
analysis is pushed to the smallest scales shown.  The difference is
particularly large for $n_s$, where the error bar is reduced by a
factor of 5.  Another parameter whose behavior is simple to understand
is $\lses$.  As discussed above, $D_{\lses}(k)$ is smaller for
\pln\ than for \pd, at all $k$.  Looking at the diagonal,
unmarginalized plots, this is why the error bars are degraded in
\pln\ when only linear scales are included.  However, when pushing
into translinear scales, the penalty from the decreased derivative
term is quickly overcome because of drastically reduced cosmic
variance, resulting in tighter constraints from \pln\ at sufficiently
small scales.

Fig.\ \ref{fig:ellipses_pvpg} shows the same figure for \pds\ and \pg.
Except for the case of the tilt $n_s$, the constraints from \pg\ are
weaker than from \pln, and often even weaker than from \pd.  This
could be surprising given that the covariance matrix of \pg\ has the
smallest non-Gaussian component, and the highest diagonality, of any
of the power spectra considered here.  The performance of \pds\ is
also disappointing given the high diagonality of its covariance
matrix; the performance is also degraded for \pds\ compared to
\pd\ for some parameters.  For \pg\ and \pds, this behavior is from
small derivative terms $\bD(k)$.  As discussed above, this is largely
from the unit variance enforced in cell densities for these density
fields.

Other analysis procedures are certainly possible.  It would be
convenient to use \pd\ on linear scales, but on translinear scales, to
exploit the reduced nonlinearity in the shape and covariance of \pg.
In principle, this could be done by setting the variance of the
Gaussian in \pg\ so that the large-scale amplitude of \pg\ matches
that of \pd, or equivalently multiplying \pg\ by a factor to line it
up with \pd\ on linear scales.  We experimented with this, estimating
this factor by averaging \pd$(k)/$\pg$(k)$ over a range of $k$, but
the uncertainty in this factor produced strong covariance on small
scales, comparable to that of \pd.  In fact, these experiments
partially motivated the form of the \pd\ covariance matrix found in
\citet{n11a}.

Fig.\ \ref{fig:ellipses_combined} shows the results from another
possibility, in which we analyze \pd\ and \pln\ together, and \pd\ and
\pg\ together.  Generally, the constraints from analyzing two power
spectra together are simply the minimum at each $\kmax$ of the
results from analyzing each individually.  This is unsurprising given
the high, but not total, degeneracy between the power spectra.
However, for the combination of \pd\ and \pg, there are significant
gains over analyzing each individually for some parameter
combinations, but the constraints are never better than for \pln.

\section{Discussion}
We have explored the sensitivity to cosmological parameters of the
power spectra of various transformations of the overdensity field.
\pd, the conventional power spectrum, benefits from being exactly the
linear power spectrum on linear scales.  Another benefit of \pd\ is in
the simple effects of smoothing on it.  However, on translinear scales
it suffers strong nonlinearities, both in the mean shape and in the
covariance, degrading parameter constraints.

\pln, the power spectrum of the log-density, has the most
cosmology-constraining power of any power spectrum considered here.
Typically, pushing to the smallest scales analyzed here, constraints
in marginalized and unmarginalized error bars are a factor of 2-3
smaller than for \pd.  The generality of this result suggests that it
would hold for other cosmological parameters as well.  This
improvement over \pd\ comes from the high diagonality of \pln's
covariance matrix, and from the small departures from the shape of the
linear power spectrum.  In particular, the tilt $n_s$ in the linear
power spectrum is dramatically better-preserved in \pln\ than in \pd,
as shown in its derivative terms in Fig.\ \ref{fig:dlogs}.  The log
transform reduces marginalized and unmarginalized error bars in $n_s$
by about a factor of 5.

\pg, the power spectrum of the rank-order-Gaussianized density field,
has a covariance matrix even more diagonal than \pln, and exceeds
\pln\ in cumulative signal-to-noise.  Also, it can be directly applied
in the case of significant discreteness noise \citep{nss11}.  This is
unlike \pln, although simple modifications of the log transform are
possible to handle the problem.  Unfortunately for cosmological
constraints, though, Gaussianization as implemented here enforces a
unit variance in cell densities.  This degrades parameter constraints,
in some cases to levels even worse than for \pd.  A notable exception
is $n_s$, for which constraints similar to \pln\ are obtained.

A promising approach explored by \citet{jtk11} employs a Box-Cox
transformation, which is a generalization of the logarithmic transform
that can be calibrated to give a distribution with vanishing skewness
and kurtosis.  Perhaps this approach can reduce the non-Gaussian
covariance to a level similar to the Gaussianization transform, while
retaining some information about the power spectrum (e.g.\ its
amplitude) on linear scales, as in the logarithmic transform.

\pds, the power spectrum of the ratio $\delta/\sigma_{\rm cell}$, where
$\sigma_{\rm cell}$ is $\delta$'s dispersion in few-Mpc cells, is the
final power spectrum that we investigate.  \pds\ has an impressively
low non-Gaussian covariance, nearly to the level of \pg.  However, in
a similar way as for \pg, dividing by the dispersion erases much of
the sensitivity to cosmological parameters, providing error bars
similar to \pd.

\section{Conclusion}
We find that applying a nonlinear transform to the nonlinear density
field can significantly enhance the cosmology-constraining power of
the power spectrum, but apparently only if the transform preserves
some linear-scale amplitude information.  The log transform, for
example, reduces error bars by a factor of 2-3; for the tilt, this
factor reaches up to 5.  The dramatic reduction in nonlinearities in
both the power spectrum covariance (as quantified previously by the
cumulative signal-to-noise ratio), and in the power spectrum shape, is
what accomplishes this.

In Paper II, we showed that issues from galaxy discreteness, perhaps
the most obvious problem for a logarithmic transform, can be overcome.
A modified logarithmic transform still enhances the cumulative
signal-to-noise ratio in the presence of discreteness noise.  If the
galaxy sampling is sufficiently dense, the tightening in parameter
constraints found in the current paper will hold when applied to
observations.

However, more work is required to investigate the cosmology sensitivity
of power spectra of Gaussianized power spectra in the face of
redshift-space distortions and galaxy bias.  In Paper II, we began
this study, but more work is required.  Generally, fingers of God,
present in redshift space, smear the density field, reducing the
non-Gaussianity of the 1-point PDF, and thus somewhat decreasing the
gains produced by a Gaussianizing transform.

Another important issue is whether Gaussianizing transforms are of use
in detecting BAO.  BAO scales are only barely translinear
(transtranslinear?), even at $z=0$, so we expect the gains from
covariance reduction alone to be modest.  As we show qualitatively in
Fig.\ \ref{fig:compline}, Gaussianizing transforms do not seem to
alter the BAO wiggles substantially, with high-order wiggles erased
similarly as in \pd\ by large-scale bulk flows.  However, the
smallest-scale wiggle or two that are not washed out lie in the regime
where the shot-noise-like one-halo term is significant, i.e.\ on the
upward ramp in the nonlinear transfer function for \pd.  This suggests
that detecting the smallest-scale wiggles may be easier for \pln\ and
\pg.

As one might expect, there are some situations in which a log
transform could be only marginally useful, and some in which it helps
substantially.  We are not aware of a case in which the transform
degrades constraints, if the analysis is pushed to sufficiently small
scales.  But even if there is such a case, analyzing the conventional
\pd\ together with \pln\ or would give tighter constraints (if only
marginally) than \pd\ alone.  Given the simplicity of these
transforms, it seems to be well-worth using them observationally.
This is even in cases that we have not directly tested, such as the
highly nonlinear scales of the galaxy power spectrum or correlation
function, sensitive to galaxy-formation details.  In this case, one
could also try looking at the ratio \pd/\pln\ (or even the ratio of
the corresponding correlation functions).

\acknowledgments I thank Istv\'{a}n Szapudi and Alex Szalay for
valuable discussions, and Katrin Heitmann and Adrian Pope for help
accessing the Coyote Universe simulations.  I also thank the anonymous
referee for useful and informative suggestions.  I am grateful for
support from the W.M.\ Keck and the Gordon and
Betty Moore Foundations, through Alex Szalay.

\bibliographystyle{hapj}
\bibliography{refs}

\end{document}